\documentclass[twocolumn,prl,floatfix,superscriptaddress]{revtex4}

\usepackage{verbatim}
\usepackage{amsmath}
\usepackage{amssymb}
\usepackage{graphicx}
\usepackage{hyperref}
\usepackage{multirow}
\usepackage{array}
\usepackage{float}
\usepackage{color}

\begin{document}
\title{Exponential Protection of Zero Modes in Majorana Islands}

\author{S.~M.~Albrecht}
\thanks{These authors contributed equally to this work.}
\affiliation{Center for Quantum Devices and Station Q Copenhagen, Niels Bohr Institute, University of Copenhagen, Copenhagen, Denmark}
\author{A.~P.~Higginbotham}
\thanks{These authors contributed equally to this work.}
\affiliation{Center for Quantum Devices and Station Q Copenhagen, Niels Bohr Institute, University of Copenhagen, Copenhagen, Denmark}
\affiliation{Department of Physics, Harvard University, Cambridge, Massachusetts, USA}
\author{M.~Madsen}
\affiliation{Center for Quantum Devices and Station Q Copenhagen, Niels Bohr Institute, University of Copenhagen, Copenhagen, Denmark}
\author{F.~Kuemmeth}
\affiliation{Center for Quantum Devices and Station Q Copenhagen, Niels Bohr Institute, University of Copenhagen, Copenhagen, Denmark}
\author{T.~S.~Jespersen}
\affiliation{Center for Quantum Devices and Station Q Copenhagen, Niels Bohr Institute, University of Copenhagen, Copenhagen, Denmark}
\author{J.~Nyg{\aa}rd}
\affiliation{Center for Quantum Devices and Station Q Copenhagen, Niels Bohr Institute, University of Copenhagen, Copenhagen, Denmark}
\author{P.~Krogstrup}
\affiliation{Center for Quantum Devices and Station Q Copenhagen, Niels Bohr Institute, University of Copenhagen, Copenhagen, Denmark}
\author{C.~M.~Marcus}
\affiliation{Center for Quantum Devices and Station Q Copenhagen, Niels Bohr Institute, University of Copenhagen, Copenhagen, Denmark}

\date{\today}

\maketitle

\textbf{Majorana zero modes are quasiparticle excitations in condensed matter systems that have been proposed as building blocks of fault-tolerant quantum computers \cite{Kitaev:2003jw}. They are expected to exhibit non-Abelian particle statistics, in contrast to the usual statistics of fermions and bosons, enabling quantum operations to be performed by braiding isolated modes around one another \cite{Kitaev:2003jw,Nayak:2008dp}. Quantum braiding operations are topologically protected insofar as these modes are pinned near zero energy, and the pinning is predicted to be exponential as the modes become spatially separated \cite{Read:392241,Kitaev:2001kla}. Following theoretical proposals \cite{Lutchyn:2010hpa,Oreg:2010gk}, several experiments have identified signatures of Majorana modes in proximitized nanowires \cite{Mourik:2012je,Rokhinson:2012ep,Das:2012hi,Deng:2012gn,Churchill:2013cq} and atomic chains \cite{NadjPerge:2014ey}, with small mode-splitting potentially explained by hybridization of Majoranas \cite{DasSarma:2012kt,Stanescu:2013je,Rainis:2013zxa}.
Here, we use Coulomb-blockade spectroscopy in an InAs nanowire segment with epitaxial aluminum, which forms a proximity-induced superconducting Coulomb island (a ÔMajorana islandÕ) that is isolated from normal-metal leads by tunnel barriers, to measure the splitting of near-zero-energy Majorana modes. We observe exponential suppression of energy splitting with increasing wire length. For short devices of a few hundred nanometers, sub-gap state energies oscillate as the magnetic field is varied, as is expected for hybridized Majorana modes. Splitting decreases by a factor of about ten for each half a micrometer of increased wire length. For devices longer than about one micrometer, transport in strong magnetic fields occurs through a zero-energy state that is energetically isolated from a continuum, yielding uniformly spaced Coulomb-blockade conductance peaks, consistent with teleportation via Majorana modes\cite{Fu:2010ho,Hutzen:2012gg}. 
Our results help to explain the trivial-to-topological transition in finite systems and to quantify the scaling of topological protection with end-mode separation.}

\begin{figure*}[t]
\center \label{fig1} 
\includegraphics[width=189mm]{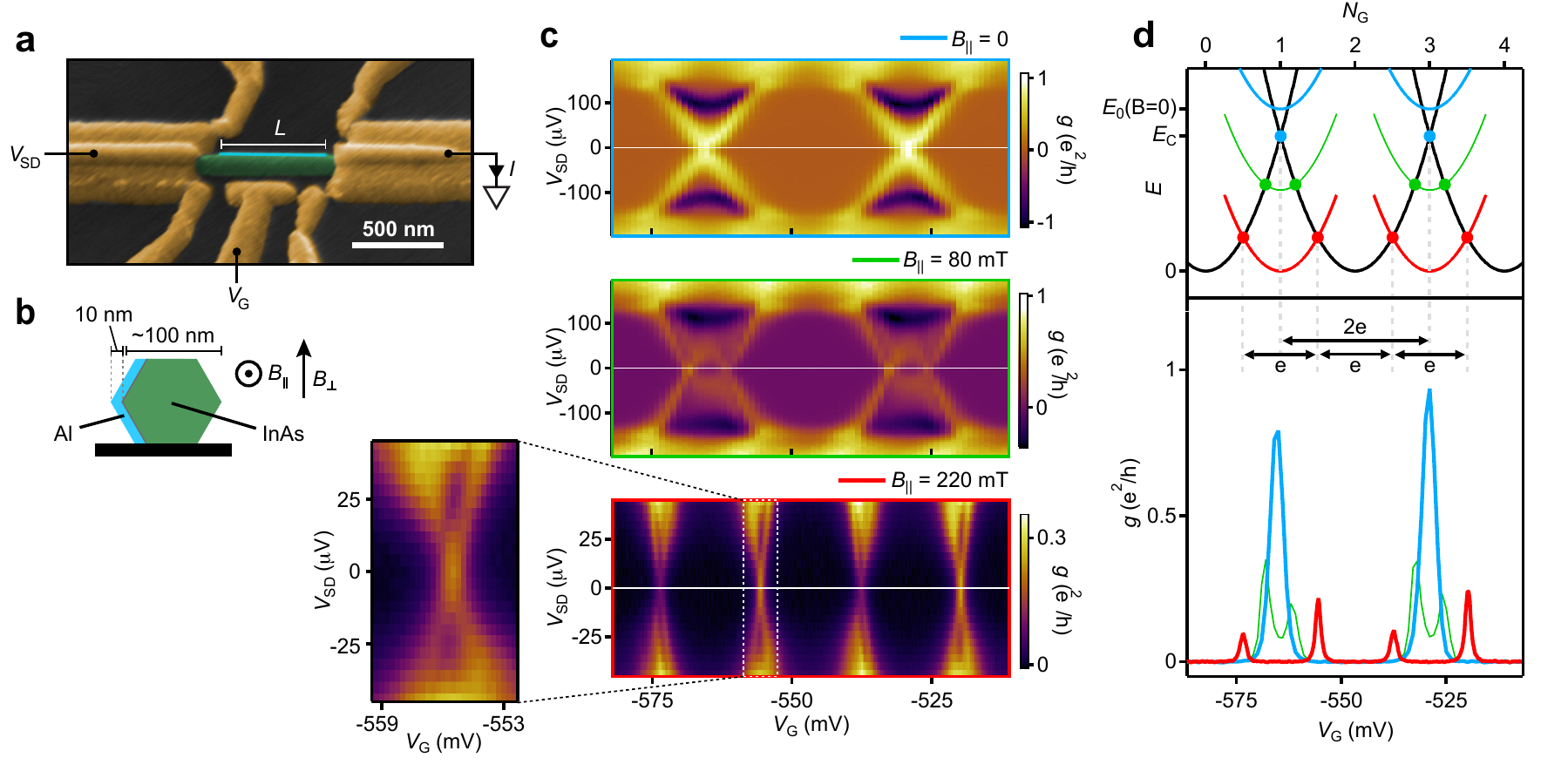}
\caption{\footnotesize{\textbf{Majorana island device.} \textbf{a},  Electron micrograph (false color) of a device that is lithographically similar to the measured devices. Gold contacts (yellow), InAs nanowire (green), and two-facet Al shell of length $L$ (light blue). Applied voltage bias, $V_ \mathrm{SD} $, and gate voltage, $V_ \mathrm{G} $, indicated. \textbf{b}, Cross section of hexagonal InAs nanowire, showing orientation of Al shell and field directions $B_\mathrm{||}$ and $B_\mathrm{\perp}$. \textbf{c}, Differential conductance, $g$, as a function of gate voltage, $V_ \mathrm{G} $, and source-drain bias, $V_\mathrm{SD} $, for parallel magnetic fields, $B_{||} = 0,80,220~\mathrm{mT}$, showing a series of Coulomb diamonds. For $B_{||} = 0$ the Coulomb diamonds are evenly spaced. An odd diamond has appeared for $B_ \mathrm{|| } = 80~\mathrm{mT}  $. For $B_\mathrm{||}  = 220~\mathrm{mT}  $ the Coulomb diamonds feature evenly spaced discrete states while the period in gate voltage has halved. Horizontal white lines indicate cuts in (d).
\textbf{d}, (upper panel) Energy $E_{N}$ of the device with electron occupancy $N$, as a function of normalized gate voltage $N_G$.
Ground-state energies for even (odd) $N$ shown in black (color).
Odd-$N$ energies are raised by the single-particle-state energy, $E_ \mathrm{0} $ compared to even-$N$ energies.
The even-$N$-only regime, $E_ \mathrm{0} > E_ \mathrm{C}$ (light blue), and the even-and-odd-$N$ regime, $E_ \mathrm{0} < E_ \mathrm{C} $ (green), where $E_\mathrm{C}$ are indicated.
The Majorana case, $E_ \mathrm{0} = 0$, is in red. Transport can occur at the intersections of parabolas, indicated by circles. (lower panel) Differential conductance, $g$, versus gate voltage $V_ \mathrm{G} $ at zero bias from measurements in (c) for magnetic fields $B_{||} = 0,80,220~\mathrm{mT}$. The splitting of the $2e$-periodic peak (light blue line) reflects a transition from Cooper pair tunneling to single-quasiparticle charging of the quantum dot.
Evenly spaced, $1e$ periodic Coulomb peaks are characteristic of a zero-energy state.}}
\end{figure*}
The set of structures we investigate consist of InAs nanowires grown by molecular beam epitaxy in the $[0001]$ wurtzite direction with an epitaxial Al shell on two facets of the hexagonal cross section \cite{2015NatMa..14..400K}. The Al shell was removed except in a small segment of length $L$ and isolated from normal metal (Ti/Au) leads by electrostatic gate-controlled barriers (Fig.~1a). Charging energy, $E_ \mathrm{C} $, of the device ranges from greater than to less than the superconducting gap of Al ($\sim$ 0.2 meV). The thin Al shell ($8-10~\mathrm{nm}$ thickness on the two facets) gives a large critical field,  $B_\mathrm{c}$, before superconductivity is destroyed: for fields along the wire axis, $B_\mathrm{c,||} \sim 1~ \mathrm{T} $; out of the plane of the substrate but roughly in the plane of the two Al-covered facets,  $B_\mathrm{c,\perp} \sim 700~ \mathrm{mT} $ (Fig.~1b). The very high achieved critical fields make these wires a suitable platform for investigating topological superconductivity \cite{2015NatMa..14..400K}.

 Five devices over a range of Al shell lengths $L \sim 0.3-1.5\,\mu$m were measured (see Methods for device layouts). Charge occupation and tunnel coupling to the leads were tuned via electrostatic gates. Differential conductance, $g$, in the Coulomb blockade regime (high-resistance barriers) was measured using standard ac lock-in techniques in a dilution refrigerator (electron temperature $\sim 50~ \mathrm{mK} $). 
 
 Figure 1c shows $g$ as a function of gate voltage, $V_\mathrm{G}$, and source-drain bias, $V_\mathrm{SD}$. For the $L = 790~\mathrm{nm}$ device, the zero-field data (top panel) show a series of evenly spaced Coulomb diamonds with a characteristic negative-differential conductance (NDC) region at higher bias. NDC is known from metallic superconductor islands \cite{Hekking:1993gc, Hergenrother:1994ei} and has recently been reported in a proximitized semiconductor device similar to those investigated here \cite{Higginbotham:2015wc}. The zero-magnetic-field diamonds reflect transport via Cooper pairs, with gate voltage period proportional to $2e$, the charge of a Cooper pair. At moderate magnetic fields (Fig.~1c, middle panel), the large diamonds shrink and a second set of diamonds appears, yielding even-odd spacing of Coulomb blockade zero-bias conductance peaks \cite{Eiles:1993uh}, as seen in the cuts in Fig.~1d. At larger magnetic fields  (Fig.~1c, lower panel) Coulomb diamonds are again periodic, now with precisely half the spacing of the zero-field diamonds, corresponding to $1e$ periodicity. NDC is absent, and resonant structure is visible within each diamond, indicating transport through discrete resonances at low bias and a continuum at high bias (see magnification in Fig.~1c). Coulomb blockade conductance peaks at high magnetic field (see Fig.~1d for zero bias cuts) with regular $1e$ periodicity (half the zero-field spacing) accompanied by a discrete subgap spectrum is a proposed signature of electron teleportation by Majorana end states \cite{Fu:2010ho,Hutzen:2012gg}. We designate as a `Majorana island' (MI) the ungrounded tunneling device in this high-field regime, where a subgap state near zero energy, energetically isolated from a continuum, leads to 1$e$-periodic Coulomb blockade conductance peaks.

Zero-bias conductance can be qualitatively understood within a simple zero-temperature model where the energy of the superconducting island---with or without subgap states (Fig.~1d)---is given by a series of shifted parabolas, $E_N( N_\mathrm{G} ) = E_ \mathrm{C}(N_ \mathrm{G} - N )^2 + p_N E_0$, where $N_ \mathrm{G}  = CV_ \mathrm{G}/e$ is the gate-induced charge (electron charge $e$ and gate capacitance $C$)  \cite{Tuominen:1992vy, Eiles:1993uh, Hekking:1993gc, Lafarge:1993zz, Hergenrother:1994ei, 1994MPLB....8.1007M}. 
$E_0$ is the energy of the lowest quasiparticle state, which is filled for odd parity ($p_{N}=1$, odd $N$), and empty for even parity ($p_N = 0$, even $N$) \cite{Higginbotham:2015wc}.
Transport occurs when the ground state has a charge degeneracy, i.e., when the $E_N$ parabolas intersect.
For $E_0 > E_\mathrm{C}$, the ground state always has even parity; transport in this regime occurs via tunneling of Cooper pairs at degeneracies of the even-$N$ parabolas. This is the regime of the $2e$-periodic Coulomb blockade peaks seen at low magnetic fields (Fig.~1d, blue).
The odd charge state is spinful and can be lowered by Zeeman energy when a magnetic field is applied. For sufficiently large field, such that $E_0 < E_\mathrm{C}$, an odd-$N$ ground state emerges. This transition from 2$e$ charging to 1$e$ charging is seen experimentally as the splitting of the $2e$-periodic Coulomb diamonds into the even-odd double-diamond pattern in Fig.~1d (green trace). In this regime the Coulomb peak spacing is proportional to $E_\mathrm{C} + 2 E_0$ for even diamonds and $E_\mathrm{C} - 2 E_0$ for odd diamonds \cite{Tuominen:1992vy, Lafarge:1993zz}.
For the particular case of a zero-energy Majorana state, $E_0 = 0$, peak spacing is regular and $1e$-periodic. This regime is observed at higher fields (Fig.~1d, red), though not sufficiently high to destroy superconductivity.

Coulomb peak spacings are measured as a function of magnetic field, allowing the state energy, $E_0( B )$, to be extracted.
An example, showing 10 consecutive peaks for the $L = 0.9~\mathrm{\mu m}$ device, is shown in Fig.~2a.
The peaks are $2e$-periodic at $B=0$, start splitting around $\sim 95~ \mathrm{mT}$, and become $1e$-periodic at $\sim110~\mathrm{mT}$, well below the spectroscopically observed closing of the superconducting gap at $B_ \mathrm{c} \sim 600~\mathrm{mT}$ (see Methods).
This points towards the presence of a state close to zero energy within the superconducting regime over a range of $\sim 500~\mathrm{mT}$.

Separately averaging even and odd Coulomb peak spacings, $\langle S_ \mathrm{e,o} \rangle$, over an ensemble of adjacent peaks reveals oscillations around the $1e$-periodic value as a function of applied magnetic field. This is consistent with an oscillating state energy $E_0$ due to hybridized Majorana modes \cite{Rainis:2013zxa,DasSarma:2012kt, Stanescu:2013je}.
For the $L = 0.9~\mathrm{\mu m}$ device (Fig.~2b), peak spacing oscillations yield an energy oscillation amplitude $A = 7.0 \pm 1.5 ~\mu \mathrm{eV}$, converted from gate voltage to energy using the gate lever arm, $\eta$, extracted independently from the slope of the Coulomb diamonds. For the $L = 1.5~\mu \mathrm{m} $ device (Fig.~2c) average Coulomb peak spacing oscillations based on 22 consecutive peaks yield a barely resolvable amplitude, $A = 1.2 \pm 0.5 ~\mu \mathrm{eV} $.

Oscillation amplitudes for the five measured devices (see Methods for device details), are shown in Fig.~2d along with a two-parameter fit to an exponential function, $ A = A_0 e^{ -L / \xi}$, giving  $A_0 = 300~ \mathrm{\mu eV} $ and $\xi = 260~ \mathrm{nm} $ as fit parameters.
The data fits well to the predicted exponential form that characterizes the topological protection of Majorana modes \cite{Read:392241, Kitaev:2001kla, DasSarma:2012kt}.  

\begin{figure}[t!]
\center \label{fig2}
\includegraphics[width=89mm]{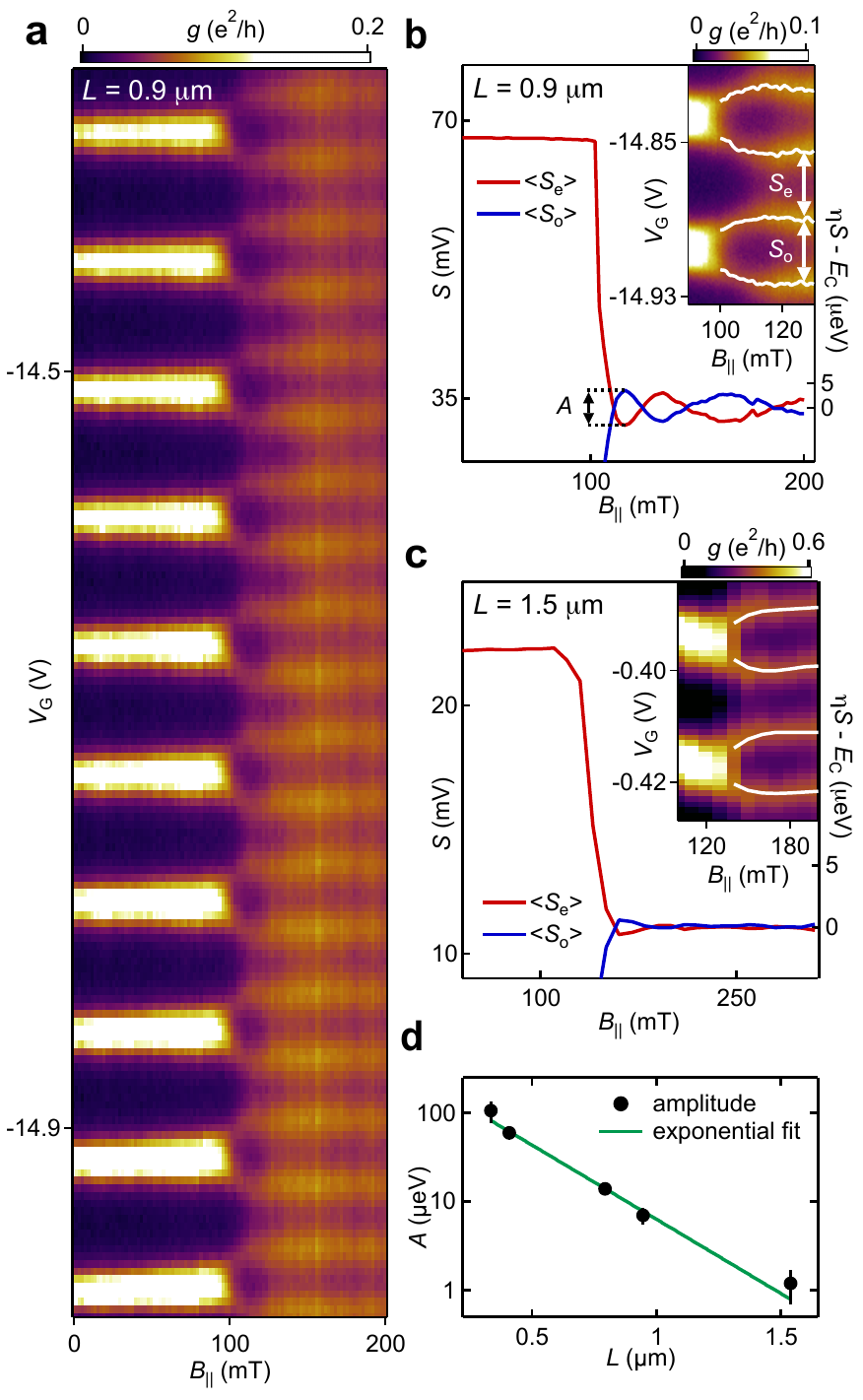}
\caption{\footnotesize{\textbf{Peak splitting in magnetic field.} \textbf{a}, Zero-bias conductance, $g$, as a function of gate voltage, $V _ \mathrm{G} $, and parallel magnetic field, $B_ \mathrm{||} $,  for $L\sim 0.9~\mathrm{\mu m} $ device, showing a series of $2e$-periodic Coulomb peaks below $\sim$\,100 mT and $1e$ nearly-periodic peaks above $\sim$\,100 mT. \textbf{b}, (inset) High-resolution measurement for $L = 0.9 ~\mu m$ (a) with overlay of peak center. Even and odd peak spacings, $S_ \mathrm{e,o} $, are indicated by arrows. (main panel) Average peak spacing for even and odd Coulomb valleys, $\langle S_ \mathrm{e,o} \rangle $, from measurement in (a) as a function of magnetic field, $B_{||}$. The Coulomb peaks become evenly spaced at $B_{||} = 110~ \mathrm{mT}$; afterwards their spacing oscillates around $\langle S_ \mathrm{e} \rangle = \langle S_ \mathrm{o} \rangle$. Right axis shows energy scale $\eta S - E_\mathrm{C} \propto E_0$ in $1e$-regime ($\eta$ is gate lever arm, see text).
\textbf{c}, Same as in (b) but for a longer wire, $L = 1.5~\mathrm{\mu m}$.
\textbf{d}, Oscillatory amplitude, $A$,  plotted against the shell length, $L$, for 5 devices from $330~\mathrm{nm} $ to $1.5~ \mathrm{ \mu m} $ (black dots) and exponential fit to $A = A_0 \exp(-L/\xi)$ with $A_0 = 300~\mathrm{\mu e V}$ and $\xi = 260~\mathrm{nm}$.
Error bars indicate uncertainties propagated from lever arm measurements and fits to peak maxima.
}}
\end{figure}

\begin{figure}[b]
\center \label{fig3}
\includegraphics[width=89mm]{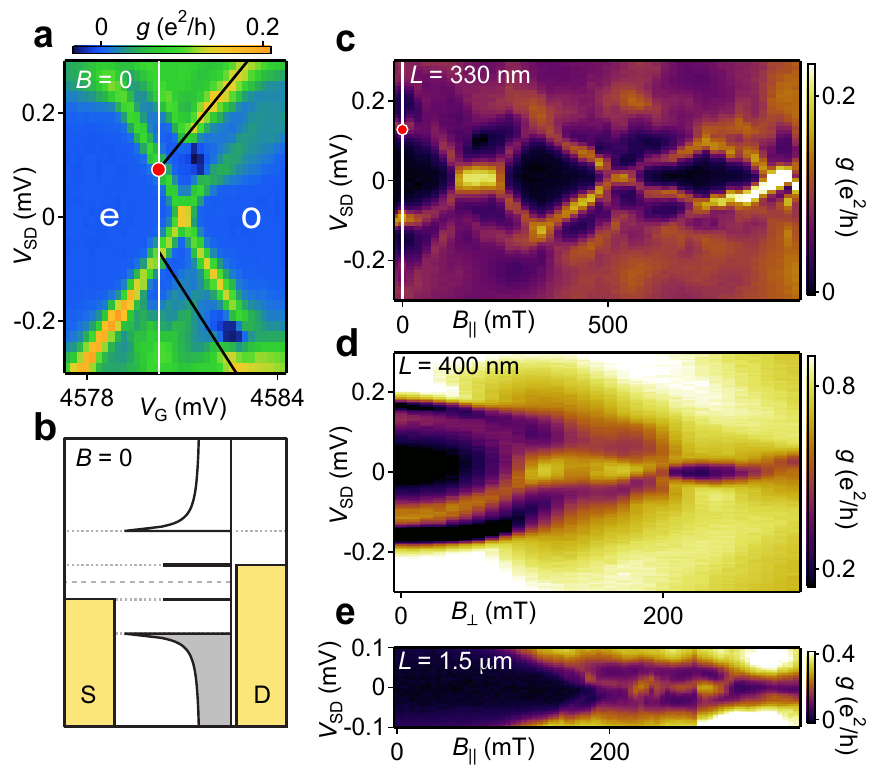}
\caption{\footnotesize{\textbf{Bias spectroscopy.}
\textbf{a}, Conductance, $g$, versus bias voltage, $V_ \mathrm{SD} $, and gate voltage, $V_ \mathrm{G} $.
Black lines indicate conductance due to bound state, red marker is at $e V_\mathrm{SD} = 2 E_0$.
\textbf{b}, Quantum dot and lead density of states at the voltage configuration indicated by red marker in (a).
Changing voltage bias moves along the white line in (a).
\textbf{c}, Conductance versus source-drain bias and magnetic field, $B_{||}$, for $L=330~\mathrm{nm}$ device with gate voltage fixed to position indicated by white line in \textbf{a}.
\textbf{d}, Conductance versus source-drain bias and magnetic field, $B_{\perp}$, for $L=400~\mathrm{nm}$ device.
\textbf{e}, Conductance versus source-drain bias and magnetic field, $B_{||}$, for $L=1.5~\mathrm{\mu m}$ device.
 }}
\end{figure}

Excited states of the MI are probed using finite-bias transport spectroscopy.
This technique requires a fixed gate voltage, chosen such that at zero bias the electrochemical potential of the leads aligns with the middle of the spectroscopic gap of the MI.
With this choice, the conductance at source-drain bias $V_\mathrm{SD}$ is due to states at energy $e V_\mathrm{SD} / 2$. A conductance peak at zero bias corresponds to a zero-energy state.
In the case shown in Figs.~3(a,b), the gate voltage is tuned using the characteristic finite-bias conductance spectra for a short InAs/Al island, investigated previously in Ref.~\cite{Higginbotham:2015wc}.
Ground-state energies determined by finite-bias spectroscopy match those extracted from zero-bias peak spacings (see Methods Fig. S7).

Bias spectroscopy shows discrete zero-energy states emerging at sufficient applied field over a range of device lengths.
In a short device (Fig.~3c), the discrete state moves linearly in magnetic field, passing through zero and merging with a continuum at $V_\mathrm{SD} \sim 100~\mathrm{\mu eV}$.
This merging is expected for Majorana systems in the short-length limit, where quenching of spin-orbit coupling results in unprotected parity crossings and state intersections at high energy \cite{Stanescu:2013je}.
Rather than passing directly through zero, the first zero crossing extends for $40~\mathrm{mT}$, which is not understood.
Medium-length devices show the subgap state bending back toward zero after zero crossings (Fig.~3d), in agreement with theoretical predictions for the emergence of Majorana behavior with increasing system length \cite{Stanescu:2013je, Rainis:2013zxa}.
For a long device ($L = 1.5\, \mu$m), bias spectroscopy shows a  zero-energy state separated from a continuum at higher bias (Fig.~3e).
The zero-energy state is present over a field range of $120~\mathrm{mT}$, with an associated energy gap $(30~\mathrm{\mu eV} ) / k_\mathrm{B} = 0.35 ~\mathrm{K}$.

The evolution from unprotected parity crossings, to energetically isolated oscillating states, and then to a fixed zero-energy state, with increasing device length is consistent with the expected crossover from a strongly overlapping precursor of split Majoranas to a topologically protected Majorana state locked at zero energy \cite{Stanescu:2013je, Rainis:2013zxa}.
Note that in the data in Fig.~3e, the signal from the discrete state disappears above $B_{||} = 320~\mathrm{mT}$. This is not expected within a simple Majorana picture. Even though the zero-bias peak disappears, the peak spacing remains $1e$-periodic (see Methods).

The observed effective g-factors, $g \sim 20-50 $, extracted from the addition spectrum and bias spectroscopy (see Methods), are large compared to previous studies on InAs nanowires \cite{Csonka:2008gz, Schroer:2011ev,Das:2012hi}, perhaps resulting from field focusing from the Al shell.
The measured gap to the continuum at zero magnetic field is consistent with the gap of aluminum $\Delta_ \mathrm{Al} \sim 180 ~\mu \mathrm{eV}  $, and is roughly the same in all devices.
The discrete subgap states (Fig.~3c-e) have zero-field energy less than but comparable to the gap, ranging from  $E_0 \left( B = 0 \right) \sim 50-160 ~\mu \mathrm{eV} $, consistent with expectations for half-shell geometries \cite{Cole:2015vf}. 
The measured gap between the near-zero-energy state and the continuum in the high-field (topological) regime, $\Delta_\mathrm{T}\sim 30~\mathrm{\mu e V}$, along with the coherence length extracted from the exponential fit to the length-dependent splitting  (Fig.~2d), $\xi \sim 260~\mathrm{nm}$, are consistent with topological superconductivity. Within this picture, at low magnetic fields, the gap and coherence length are related to the strength of spin-orbit coupling, yielding a value $\alpha_\mathrm{SO} \sim \xi \cdot \Delta_\mathrm{T}  = 8\times 10^{-2}~\mathrm{eV \cdot \AA}$ that is consistent with previously reported values in InAs nanowires \cite{Fasth:2007ie,Das:2012hi}.
For a single subband picture, this implies a Fermi velocity \mbox{$v_\mathrm{F} =\alpha_ \mathrm{SO}  / \hbar = 1 \times 10^{4}~\mathrm{m/s}$} that is lower than expected, suggesting that more than one subband is occupied under the Al shell, though we are not able to extract the number of modes directly.

\begin{figure}
\center \label{fig4}
\includegraphics[width=80mm]{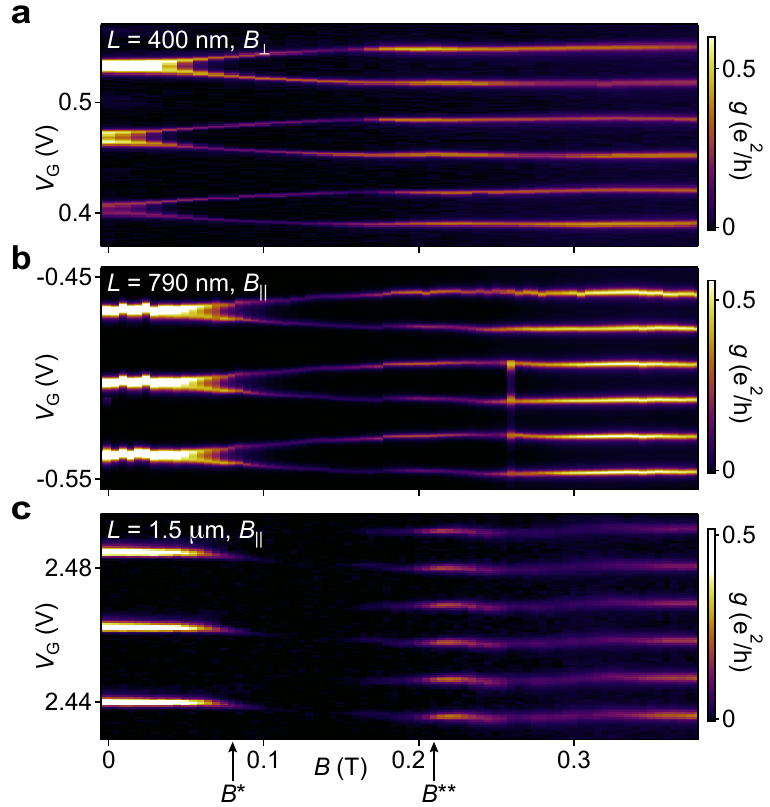}
\caption{\footnotesize{\textbf{Length dependence of Coulomb peak heights.}
\textbf{a-c}, Conductance as a function of magnetic field and gate voltage for device lengths $L = 400~\mathrm{nm}$, $790~\mathrm{nm}$, $1.5~\mathrm{\mu m}$.
Coulomb peaks become dim at field $B^{*}$ and brighten at field $B^{**}$, particularly for the $L=1.5~\mathrm{\mu m}$ device, consistent with teleportation at fields above above $B^{**}$.
}}
\end{figure}

Finally, we consider the magnetic field dependence of Coulomb blockade peak heights (as opposed to spacings), as seen in Fig.~4. We found in most devices that below the field $B^{*}$ where $2e$-periodic peaks split, all peaks were uniformly high amplitude. Above $B^{*}$, peak heights were rapidly suppressed and remained low up to a second characteristic field, $B^{**}$, coincident with $1e$ periodicity (i.e., the field where  even-odd spacing differences vanished). Above $B^{**}$, peak heights recovered. In the longer wires, peaks were nearly absent between $B^{*}$ and $B^{**}$, as seen in Fig. 4c. 

We interpret these observations as follows: In the present lead-wire-lead geometry, transport above $B^{*}$ involves single electrons entering one end of the wire and leaving from the other.  The onset of uniform spacing with the reappearance of high peaks above $B^{**}$  indicates the emergence of a state (or states) at zero energy with strong wave function support at both ends of the wire. This is consistent with teleportation of electrons from one end of the wire to the other via a Majorana mode \cite {Fu:2010ho,Hutzen:2012gg}, though not necessarily a unique signature \cite{2015PhRvB..92b0511S}. Thus while the simultaneous brightening of peaks with their becoming uniformly spaced at $B^{**}$ suggests a subgap/Majorana mode moving to the ends of the wire as it moves to zero energy, we cannot rule out other forms of end-localized zero-energy states that could appear above a critical field.

In summary, we have studied Majorana islands composed of InAs nanowires covered on two facets with epitaxial Al, for a range of device lengths. 
Zero-energy states are observed for wires of all lengths away from zero field. Oscillating energy splittings, measured using Coulomb blockade spectroscopy, are exponentially suppressed with wire length, with a characteristic length $\xi = 260~\mathrm{nm}$. This constitutes an explicit demonstration of exponential protection of zero-energy modes.
Finite-bias measurements show transport through a discrete zero-energy state, with a measured topological gap $\Delta_\mathrm{T} = 30~\mathrm{\mu eV}$ for long devices.
The extracted $\Delta_\mathrm{T}$ and $\xi$ are consistent with known parameters for InAs nanowires and the emergence of topological superconductivity.
Brightening of Coulomb peaks at the field where spacing becomes uniform for longer devices suggests the presence of a robust delocalized state connecting the leads, and provides experimental support for  electron teleportation via Majorana modes.

\vspace{0.1in}

\textbf{Acknowledgements}

We thank K.~Flensberg, M.~Leijnse, M.~Deng, W.~Chang, and R.~Lutchyn for valuable discussions, and G.~Ungaretti, S.~Upadhyay, C.~S{\o}rensen, M.~von Soosten, and D.~Sherman for contributions to growth and fabrication. Research supported by Microsoft Project Q, the Danish National Research Foundation, the Lundbeck Foundation, the Carlsberg Foundation, and the European Commission. CMM acknowledges support from the Villum Foundation.

\vspace{0.1in}

\textbf{Author contributions}

P.K., T.S.J. and J.N. developed the nanowire materials. S.M.A. fabricated the devices. S.M.A., A.P.H. and M.M. carried out the measurements with input from F.K., T.S.J. and C.M.M. Data analysis was done by S.M.A., A.P.H. and M.M. All authors contributed to interpreting the data. The manuscript was written by S.M.A., A.P.H. and C.M.M. with suggestions from all other authors.

\vspace{0.1in}


\begin{widetext}
\section{Methods}
\setcounter{figure}{0}
\renewcommand{\figurename}{EXTENDED DATA FIG.}
\renewcommand{\tablename}{EXTENDED DATA TABLE}
\renewcommand{\thetable}{\arabic{table}}  

\subsection{Sample preparation}
The InAs nanowires with epitaxial Al shell were grown via a two-step process by molecular beam epitaxy. First, the InAs nanowires were grown using the vapor-liquid-solid method with Au as a catalyst at $420^\circ$C. Second, after cooling the system to $-30^\circ$C the Al was grown on two facets of the hexagonal cross section \cite{2015NatMa..14..400K}. Afterwards the nanowires were deposited on degenerately doped Si substrates with 100-500 nm thick thermal oxides using either wet or dry deposition techniques. Wet deposition involves sonicating a growth substrate of nanowires in methanol for a few seconds, then putting several drops of the nanowire-methanol solution onto the chip surface using a pipette. Dry deposition was done by bringing a small piece of cleanroom wipe in touch with the growth substrate, then afterwards swiping it onto the chip surface. We find that while wet deposition results in a more uniform dispersion of nanowires on the chip surface, dry deposition is faster and less wasteful with nanowires. Selective removal of the Al shell was done by patterning etch windows using electron beam lithography on both sides of the nanowire, plasma cleaning the surface of the nanowire using oxygen, then etching the Al using a Transene Al Etchant D with an etching time of 10 seconds at $50^\circ$C. Depending on the device, ohmic contacts to the InAs core were fabricated using either ion milling or sulfur passivation to remove surface oxides. Ion milling was done for times ranging from 85 s to 110 s using a Kaufman \& Robinson KDC 40 4-CM DC Ion Source with an acceleration voltage of 120V and an ion beam current density of 0.5 mA/$\mathrm{cm^2}$ at the chip surface. Sulfur passivation was done using a 2.1\% solution of $\mathrm{(NH_4)_2S}$ in DI water with 0.15 M dissolved elemental sulfur at $40^\circ$C for 20 minutes. This was followed by the deposition of $5~ \mathrm{nm}$ of Ti as a sticking layer and $ 70-100~ \mathrm{nm} $ of Au for the ohmic contact. We found that ion milling resulted in more stable devices. Side and plunger gates were lithographically defined in the same fabrication step as the ohmic contacts in order to increase device yield. PMMA was used as resist in all lithography steps.

\subsection{Device Geometries}
Gate patterns of the five measured devices are shown in ED Fig.~\ref{figgeo}. With the exception of the $L = 0.9 ~ \mathrm{\mu m} $ device, all measurements involving gate dependence are tuned through resonances using the plunger gate on either the Al side or the uncoated InAs side. For the $ L = 0.9 ~ \mathrm{\mu m} $ device, the lower left side gate is used to tune through resonances of the quantum dot, because the central plunger gate was not bonded during the cool down.
\begin{figure*}
\center
\includegraphics[width=183mm]{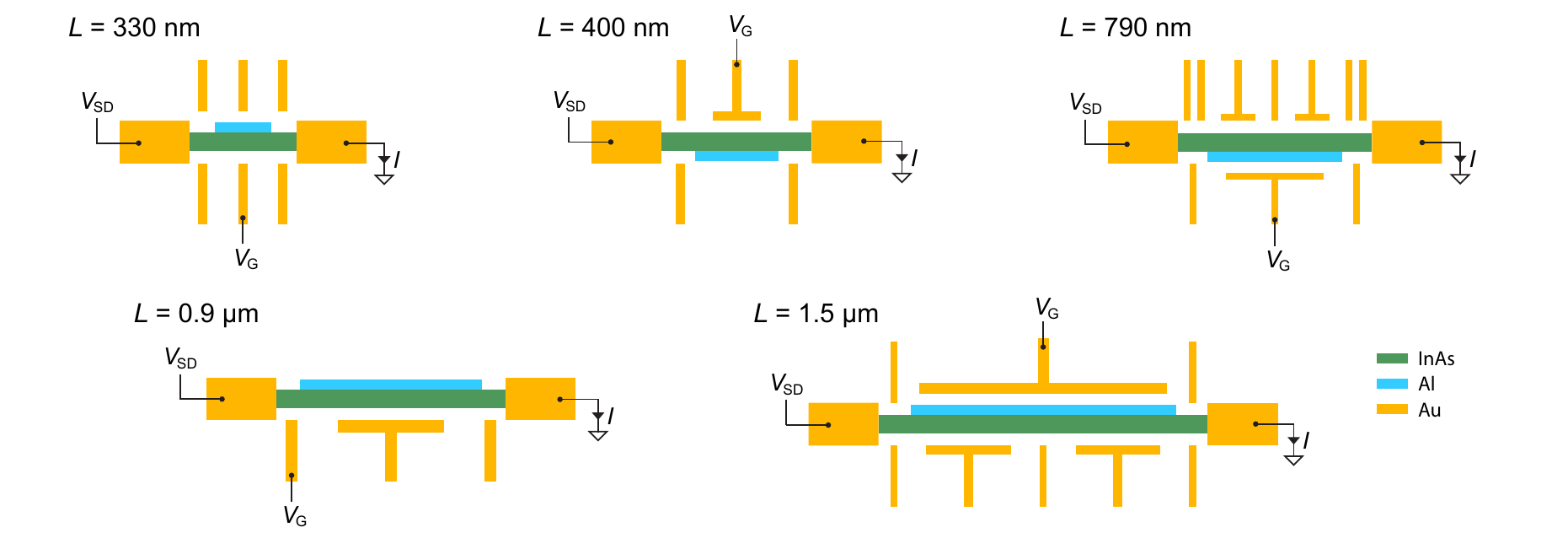}
\caption{\footnotesize{\textbf{Device Layouts.} Gate pattern for the five measured devices showing applied voltage bias, $V_ \mathrm{SD} $, and gate voltage, $V_ \mathrm{G} $.}}
\label{figgeo}
\end{figure*}

\subsection{Measurements}
Transport measurements were carried out in an Oxford Triton dilution refrigerator with a base electron temperature of $T \sim 50~\mathrm{mK} $ and a 6-1-1 T vector magnet. Differential conductance, $g = dI/dV_\mathrm{SD}  $, was measured using the AC-lockin technique with an excitation voltage in the range 2-6 $\mathrm{\mu V} $.

\subsection{Peak spacing data summary}

\begin{figure*}[h!]
\center
\includegraphics[width=160mm]{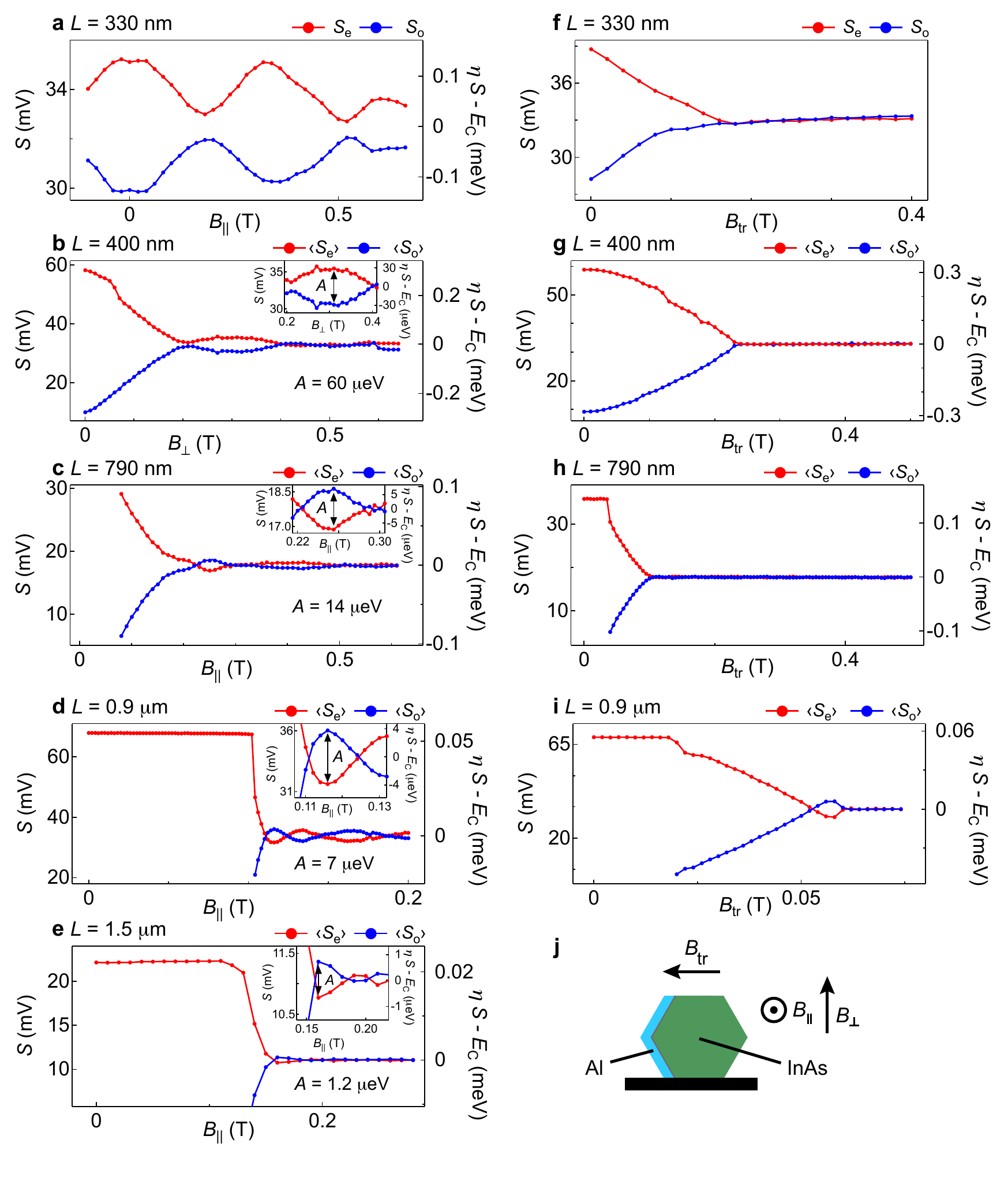}
\caption{\footnotesize{\textbf{Even-odd peak spacing summary.} \textbf{a-i}, Peak spacings for even and odd valleys, $S_ \mathrm{e,o} $, versus applied magnetic field, similar to Fig. 2b, for different device lengths. Left axis shows peak spacings, right axis shows corresponding energy scales, converting from gate voltage to energy by the lever arm, $\eta$, measured independently from Coulomb blockade diamonds. Inset shows a magnification of the first energy splitting with an arrow indicating where $A$ is measured. \textbf{j}, Cross section of the nanowire, showing the applied field directions $B_{||}$, $B_\perp$ and $B_\mathrm{tr}$.}
\label{fig:eoSummary}}
\end{figure*}

The exponential curve in Fig.~2d (main text) is derived from even-odd peak spacing measurements in the high critical field directions, $B_{||}$ and $B_\perp$, summarized in ED Fig.~\ref{fig:eoSummary}.
Suppression of spacing fluctuations with increased device length is clearly visible.
The measured $A$ is indicated by black arrows in the inset, and the values are recorded in ED Table~\ref{tab:Ecsummary} for each device length, along with charging energies and lever arms.

For $L=330~\mathrm{nm}$, Coulomb peak fluctuations became uncorrelated after several peaks.
To obtain a large statistical ensemble, fluctuations were averaged over five sets of Coulomb peaks taken in different device tunings.
ED Fig.~\ref{fig:eoSummary}a shows data from a single set of peaks, and ED Table~\ref{tab:Ecsummary} reports the full ensemble average.

In a transverse magnetic field applied in the low critical field direction, $B_\mathrm{tr}$, shown in ED Fig.~\ref{fig:eoSummary}f-i, the oscillations are absent, with the exception of an initial overshoot for $L = 0.9~\mathrm{\mu m}$ at $B_\mathrm{tr} = 55~\mathrm{mT}$ (ED Fig.~\ref{fig:eoSummary}i), before the system is driven into the normal state at $B_\mathrm{tr} \sim 65~\mathrm{mT}$.

\begin{table}
\centering
\begin{tabular}{c  c  c  c}
L [$\mathrm{nm}$]\ \ & $E_\mathrm{C}$ [$\mathrm{meV}$]\ \ & $\eta$ [$\mathrm{eV/V}$]\ \ & $A$ [$\mu \mathrm{eV}$]\ \  \\ 
\hline
330 & 1.6 & 0.048 & 106\\
400 & 0.40 & 0.012 & 60\\
790 &  0.14 & 0.008 & 14\\
950 & 0.054 & 0.0016 & 7\\
1540 & 0.022 & 0.002 & 1.2\\
\end{tabular}
\caption{ \label{tab:Ecsummary}
Device length, $L$, charging energy, $E_\mathrm{C}$, lever arm, $\eta$, and characteristic amplitude, $A$, for the five measured devices.
}
\end{table}
\begin{figure}[t]
\center 
\includegraphics[width=89mm]{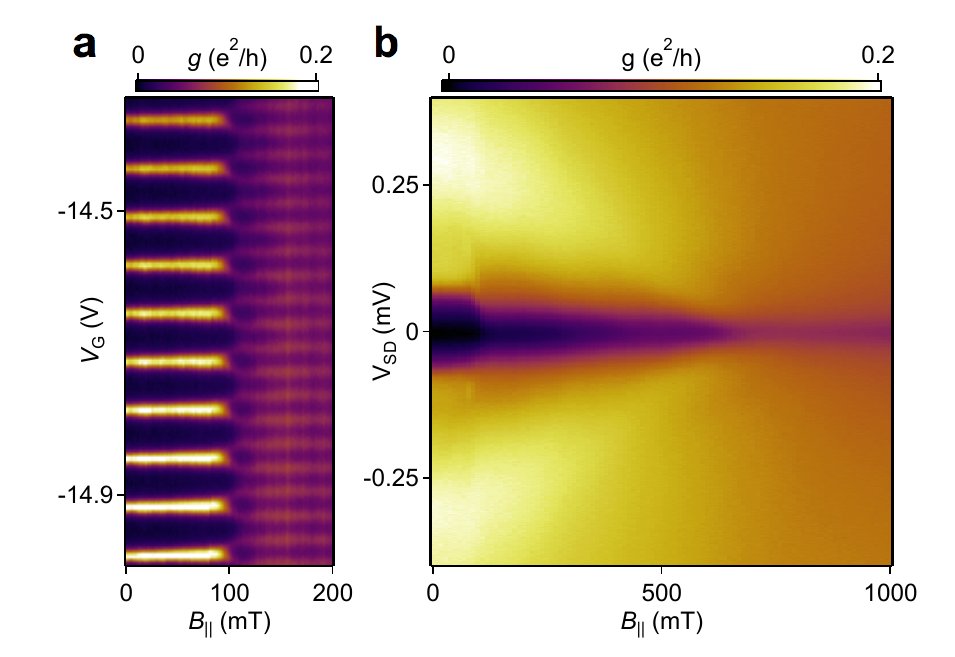}
\caption{\footnotesize{\textbf{Critical field measurement for 0.9~$\mathrm{\mu}$m device.} \textbf{a}, Conductance, $g$, versus gate voltage, $V_ \mathrm{G}$, and parallel magnetic field, $B_{||}$, at zero bias showing the $2e$-$1e$ peak splitting. \textbf{b}, Conductance versus source drain voltage, $V_ \mathrm{SD} $, and $B_{||}$, taken at $V_ \mathrm{G}= -14.92~ \mathrm{V}  $, showing a closing of the superconducting gap at $B_c \sim 640 ~ \mathrm{mT} $, more than $500~\mathrm{mT} $ after the onset of $1e$-periodicity.}\label{figBC}}
\end{figure}

\begin{figure*}
\center 
\includegraphics[width=150mm]{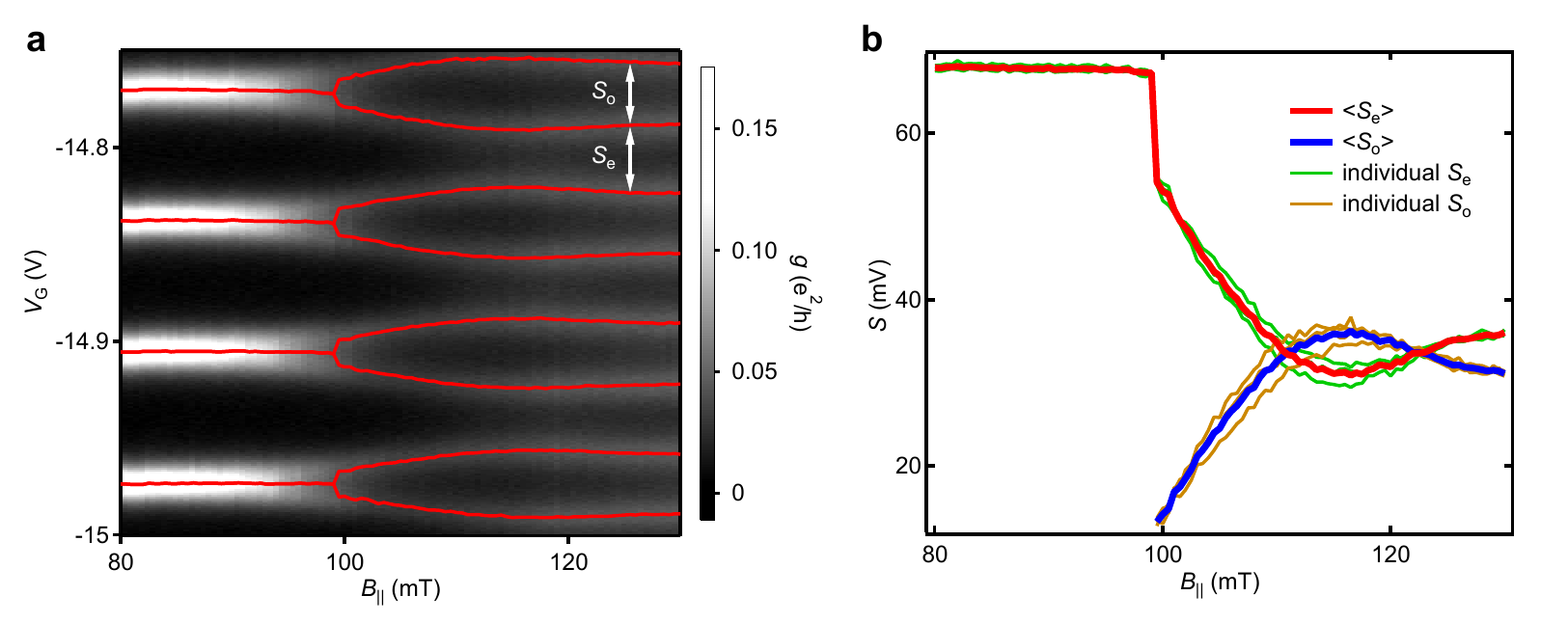}
\caption{\footnotesize{\textbf{Oscillating $\bf 1e$-periodic peak spacings.} \textbf{a}, Zero-bias conductance, $g$, versus gate voltage, $V_ \mathrm{G}$, and parallel magnetic field, $B_{||}$, at zero bias showing the $2e$-$1e$ peak splitting for $L = 0.9~\mathrm{\mu m}$. The fit peak position is indicated by a red line, even and odd peak spacings, $S_ \mathrm{e,o} $, indicated by white arrows. \textbf{b}, Peak spacing, $S_ \mathrm{e,o}$, for even and odd valleys as a function of $B_{||}$. The plot shows the average peak spacing $\langle S_ \mathrm{e,o} \rangle $ as well as the individual peak spacings, $S_ \mathrm{e,o} $.}\label{figpeaks}}
\end{figure*}

\begin{figure*}[h]
\center
\includegraphics[width=183mm]{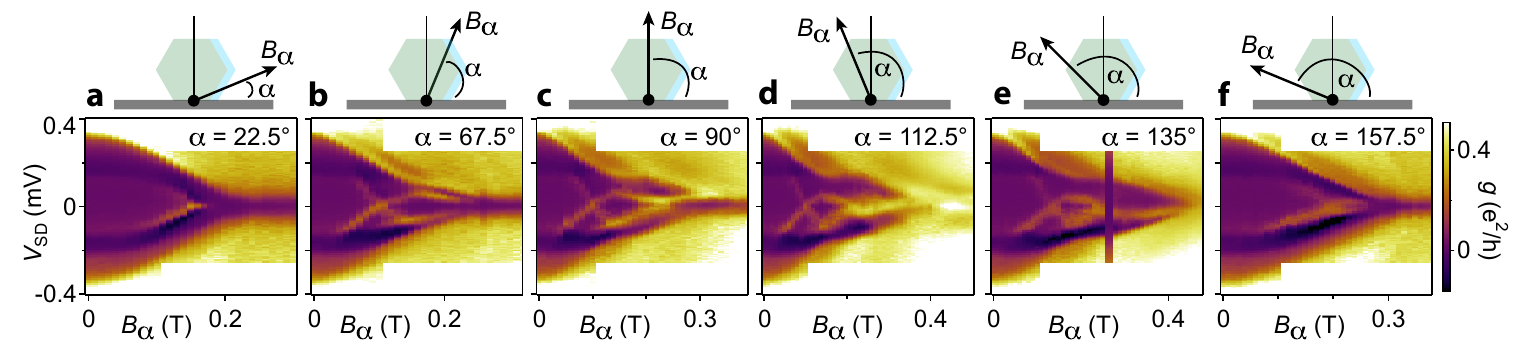}
\caption{\footnotesize{\textbf{Angle dependence of state-continuum anti-crossing.} \textbf{a-f}, Differential conductance, $g$, as a function of source-drain bias, $V_ \mathrm{SD} $, and magnetic field $B_\alpha $ for different angles, $\alpha = 22.5-157.5^\circ$, in the plane perpendicular to the nanowire direction. Measurements are from the $L = 400~\mathrm{nm}$ device.}}
\label{figangle}
\end{figure*}

\subsection{Magnetic Field Orientation}
The direction of the nanowire on the chip was found by orienting the magnetic field from a vector magnet in the chip plane and spectroscopically measuring the anisotropy of the critical magnetic field. By comparing to the wire-direction based on optical and electron micrographs, we estimate an angular precision of $\pm ~3$ degrees.

\subsection{Critical Field Measurements}
The observed $2e$-$1e$ splitting at $B_{||} \sim 95~ \mathrm{mT} $ is compared to the closing of the superconducting gap at a considerably higher critical field, $B_{c,||}$, in ED Fig.~\ref{figBC}. Bias spectroscopy in ED Fig.~\ref{figBC}b shows a closing of the superconducting gap at $B_{c,||} \sim 600~\mathrm{mT}$, more than 500~mT after the onset of evenly spaced $1e$-periodic Coulomb peaks. The change from $2e$ to $1e$-periodicity at $B_{||} \sim 100~ \mathrm{mT} $ in ED Fig.~\ref{figBC}a coincides with a reduction in the measured Coulomb gap in ED Fig.~\ref{figBC}b, reflecting the transition from Cooper pair charging (energy penalty $2E_ \mathrm{C} $) to single-electron charging (energy penalty $E_ \mathrm{C} $). The measurement in ED Fig.~\ref{figBC}b was taken in a Coulomb valley at the gate voltage $V_ \mathrm{G}= -14.92~ \mathrm{V}$.

\subsection{Averaging of peak spacings}
In Fig.~2b of the main text, we show the extracted average peak spacing for several even and odd Coulomb valleys. A high resolution measurement of the $2e$-$1e$ splitting is shown in ED Fig.~\ref{figpeaks}a. The individual even and odd valleys, $S_ \mathrm{e,o} $ in ED Fig.~\ref{figpeaks}b, exhibit the same oscillating behavior but show a small deviation from the average between $100-125~ \mathrm{mT} $, which might be attributable to $g$-factor fluctuations for successive charge occupations of the quantum dot.  Below 100 mT the fluctuations are very small, giving an indication of instrumental noise in the measurement.

\subsection{Angle Dependence}
Angle dependence of the anti-crossing of the state with the continuum for $L = 400~ \mathrm{nm} $ is shown in ED Fig.~\ref{figangle}. We focus on magnetic fields, $B_\alpha$, with angles, $\alpha$, in the plane perpendicular to the nanowire direction.
The measurements show a pronounced anti-crossing between the sub-gap state and an excitation continuum ($\alpha = 112.5^\circ$ and $\alpha = 135^\circ$) that is significantly reduced for $\alpha = 67.5^\circ$. Interpreting angle dependence is complicated by the anisotropy of g-factor and critical field. 
The critical field is maximized for $\alpha=120^\circ$, and is reduced drastically for near-perpendicular field alignment ($\alpha = 22.5^\circ$).
The observed g-factors are highly dependent on field orientation and device tuning.
For the $L = 400~ \mathrm{nm} $ device shown in ED Fig.~\ref{figangle}, we found an approximately sinusoidal variation in $g$-factor by a factor of 2, with maximum g-factor occurring near $\alpha = 90^\circ$.

\subsection{Choice of gate voltage for bias spectroscopy}

\begin{figure*}
\center 
\includegraphics[width=183mm]{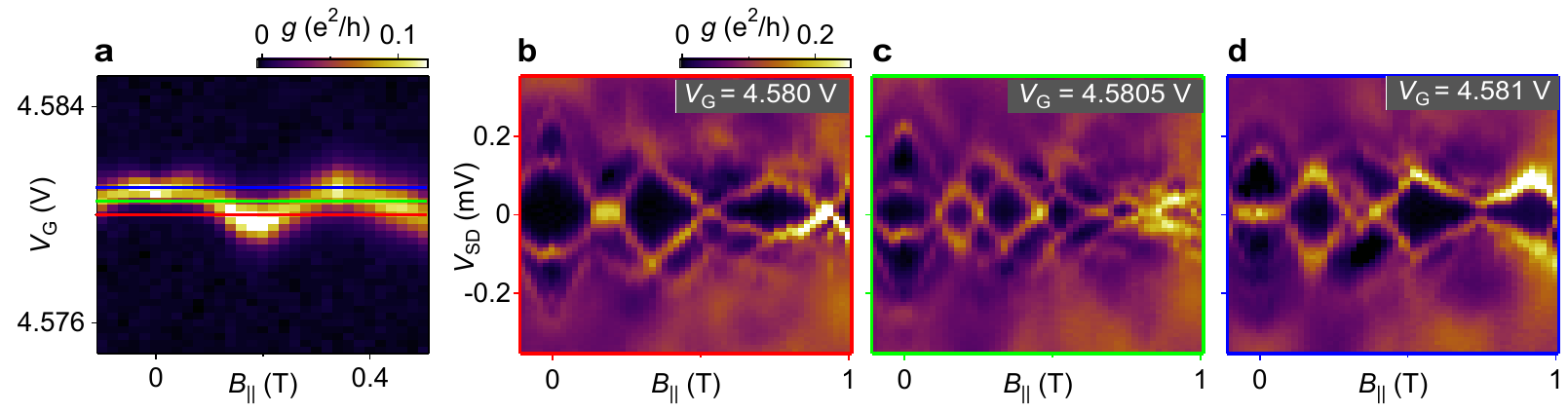}
\caption{\footnotesize{\textbf{Gate positions.} \textbf{a}, Differential conductance, $g$, as a function of gate voltage, $V_ \mathrm{G} $,  and parallel magnetic field, $B_{||}$, for $L = 330~ \mathrm{nm} $. Three different gate positions are indicated by colored horizontal lines. \textbf{b-d}, Differential conductance as function of bias voltage, $V_ \mathrm{SD} $, and $B_{||}$ for the three gate voltages in (a).}}
\label{figSilvanos}
\end{figure*}

For bias spectroscopy, the gate voltage is fixed either by interpreting Coulomb diamonds, as discussed in the main text, or from even-odd peak spacings.
While details of the bias spectroscopy, such as locations of zero-crossing, depend on the choice of gate voltage, general features such as slopes, typical fluctuation amplitude, and the presence of a robust excitation gap are not strongly affected by the choice of gate voltage (ED Fig.~\ref{figSilvanos}).

\subsection{Comparison of addition energies and finite bias spectroscopy}

\begin{figure}[h]
\center 
\includegraphics{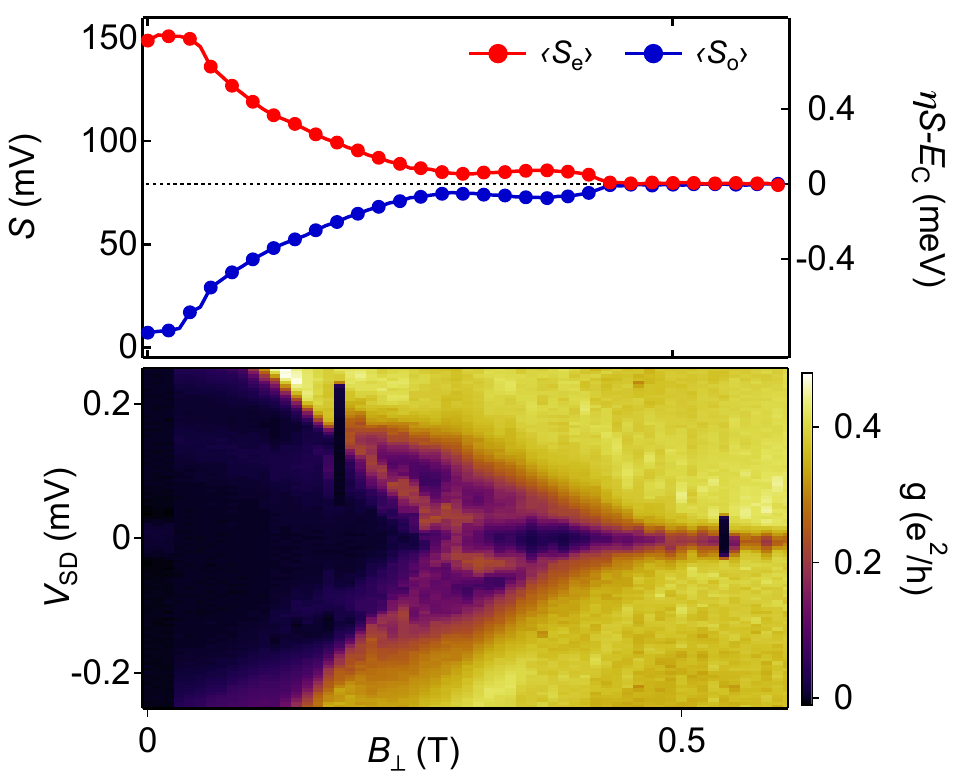}
\caption{\footnotesize{\textbf{Comparison of peak spacings and bias spectroscopy.} \textbf{a}, Peak spacing for even and odd valleys, $\left< S_{ \mathrm{e,o} } \right>$, versus applied field, $B_\perp$.
\textbf{b}, Differential conductance, $g$, as a function of source-drain bias, $V_ \mathrm{SD} $, and magnetic field.
}}
\label{figMethodComp}
\end{figure}
\begin{figure}[h]
\center 
\includegraphics{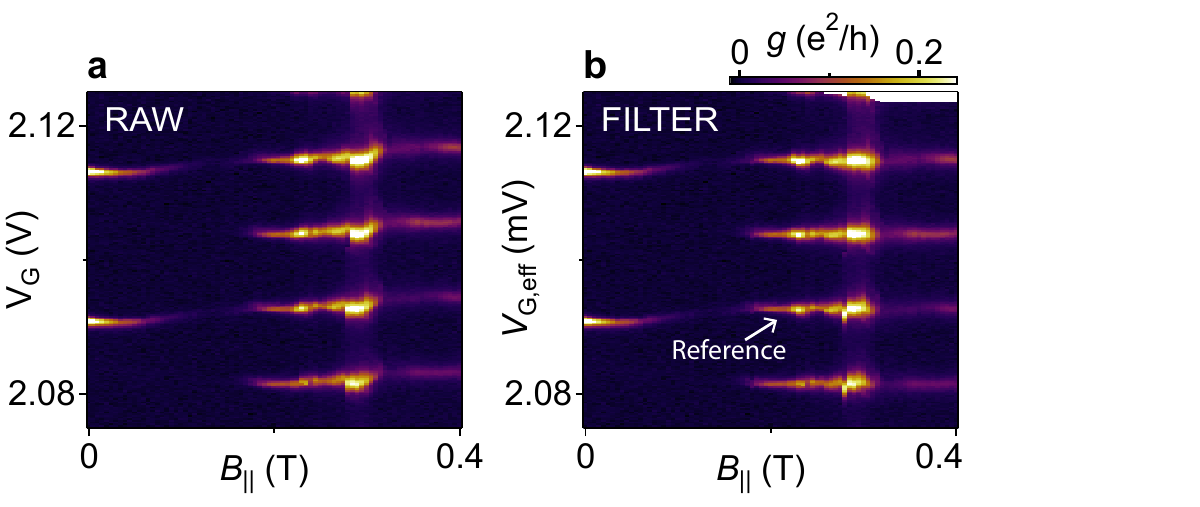}
\caption{\footnotesize{\textbf{Common-mode peak motion removal.} \textbf{a}, Differential conductance, $g$, versus gate voltage, $V_\mathrm{G}$ and applied magnetic field $B_{||}$, for $L=1.5~\mathrm{\mu m}$ device. 
\textbf{ b }, Same as (a), but with effective gate voltage, $V_\mathrm{G,eff}$, defined to remove common-mode peak motion.
Reference Coulomb peak, used for common-mode removal, labeled.
}}
\label{figCMode}
\end{figure}

Peak spacings are used to measure the energy of the lowest-lying state.
The same information is present in the bias spectroscopy, and gives consistent results, as shown in ED Fig.~\ref{figMethodComp}.

\subsection{Bias spectroscopy of long device}
\label{sec:longBias}

Common-mode fluctuations in Coulomb peak position were observed in the longest ($L=1.5~\mathrm{\mu m}$) device, as shown in ED Fig.~\ref{figCMode}a.
The fluctuations evidently correspond to a shift in the electrochemical potential of the dot, likely due to a nearby, field-dependent charge trap.
The fluctuations are small compared to charging energy, but complicate the application of bias spectroscopy which needs to be performed at fixed electrochemical potential.
To correct for the fluctuations, we introduce an effective gate voltage,
\begin{equation}
V_\mathrm{G,eff}( B ) = V_\mathrm{G} + \delta V( B ),
\end{equation}
that removes the common-mode peak motion. 
The offset voltage is zero at low field, when Coulomb peaks are $2e$ periodic ($\delta V( B ) = 0$ for $B \leq 175~\mathrm{mT}$).
At high field, $\delta V( B )$ is chosen so that the reference Coulomb peak (labeled in ED Fig.~\ref{figCMode}b) occurs at constant $V_\mathrm{G,eff}$.
All nonzero $\delta V( B )$ are listed in ED Table \ref{tab:deltaV}.

\begin{table}
\centering
\begin{tabular}{c  c}
B (mT)\ \  & $\delta V~(\mathrm{mV})$ \\
\hline
180 & 0.25 \\
230 & 0.25 \\
235 & 0.25 \\
240 & 0.25 \\
245 & 0.25 \\
250 & 0.25 \\
255 & 0.25 \\
260 & 0.5 \\
265 & 0.5 \\
270 & 0.5 \\
275 & 0.5 \\
280 & 0.75 \\
300 & 0.25 \\
305 & 0.75 \\
310 & 1.25 \\
315 & 1.5 \\
320 & 1.75 \\
325 & 1.75 \\
330 & 1.75 \\
335 & 1.75 \\
340 & 1.75 \\
345 & 1.75 \\
350 & 1.75 \\
355 & 1.75 \\
360 & 1.75 \\
365 & 1.75 \\
370 & 1.75 \\
375 & 1.75 \\
380 & 1.75 \\
385 & 1.75 \\
390 & 1.75 \\
395 & 1.75 \\
400 & 1.75 \\
\end{tabular}
\caption{ \label{tab:deltaV}
All nonzero offset voltage values, $\delta V( B )$, for $L=1.5~\mathrm{\mu m}$ device.
Offset is defined for $B=0, 5, 10, ..., 400 ~ \mathrm{mT}$.
}
\end{table}

As shown in ED Fig.~\ref{figCMode}b, this procedure removes the common-mode peak motion.
In the case of the $1.5~\mathrm{\mu m}$ device, bias spectroscopy is performed at fixed $V_\mathrm{G,eff}$, which allows us to infer the energy of the sub-gap state at fixed electrochemical potential.

\subsection{Zero-energy state at successive Coulomb peaks}

The zero-energy state is robust over many successive Coulomb peaks, as shown in ED Fig.~\ref{figDiamonds}.
The full bias spectroscopy as a function of field is also reproducible over several peaks, as shown in ED Fig.~\ref{fig4zbps}.

\begin{figure}
\center
\includegraphics[width=0.5\textwidth]{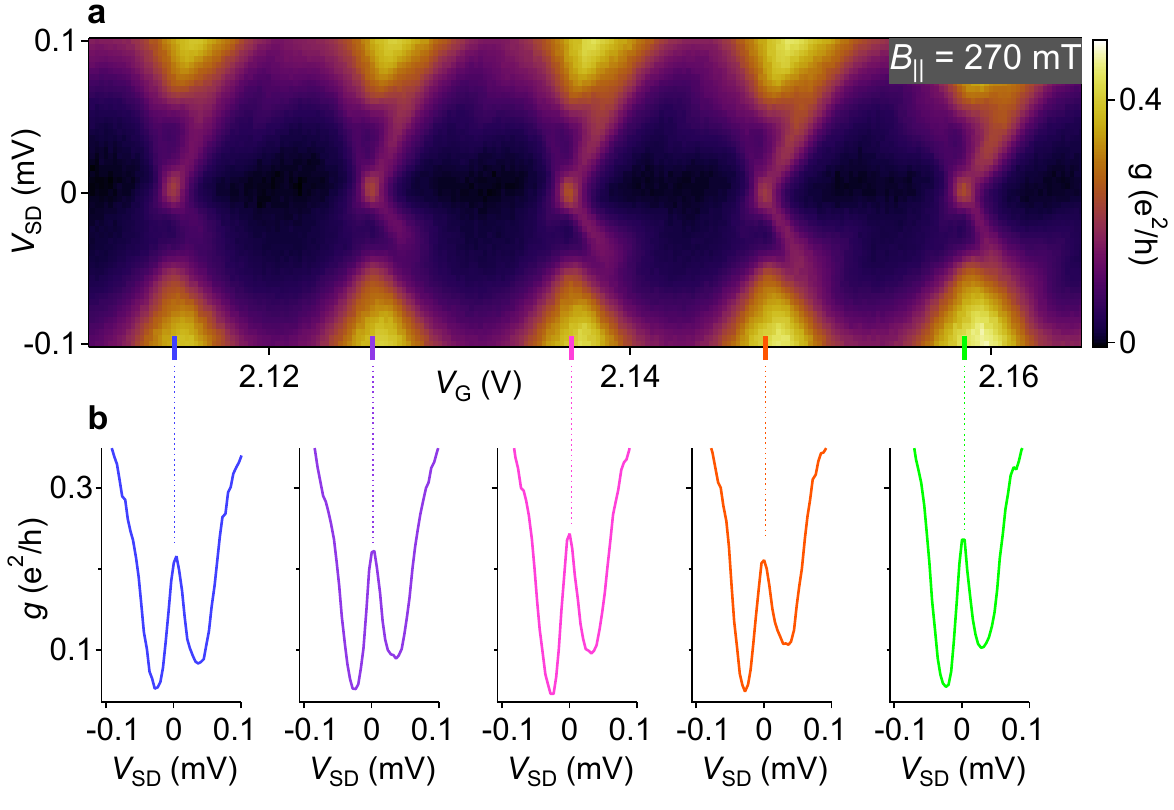}
\caption{\footnotesize{\textbf{Zero-energy state.} \textbf{a}, Differential conductance, $g$, as a function of bias voltage, $V_ \mathrm{SD} $, and gate voltage, $V_ \mathrm{G} $, for $L = 1.5~\mu \mathrm{m} $ and $B_{||} = 270~ \mathrm{mT}$, showing an evenly spaced Coulomb diamond pattern and the associated gapped zero-energy state. \textbf{b}, Differential conductance versus $V_\mathrm{SD}$, at the gate voltages indicated by colored ticks in (a). At these $V_ \mathrm{G} $ values, the presence of a zero-energy state is indicated by a zero bias peak.}}
\label{figDiamonds}
\end{figure}

\begin{figure}
\center 
\includegraphics{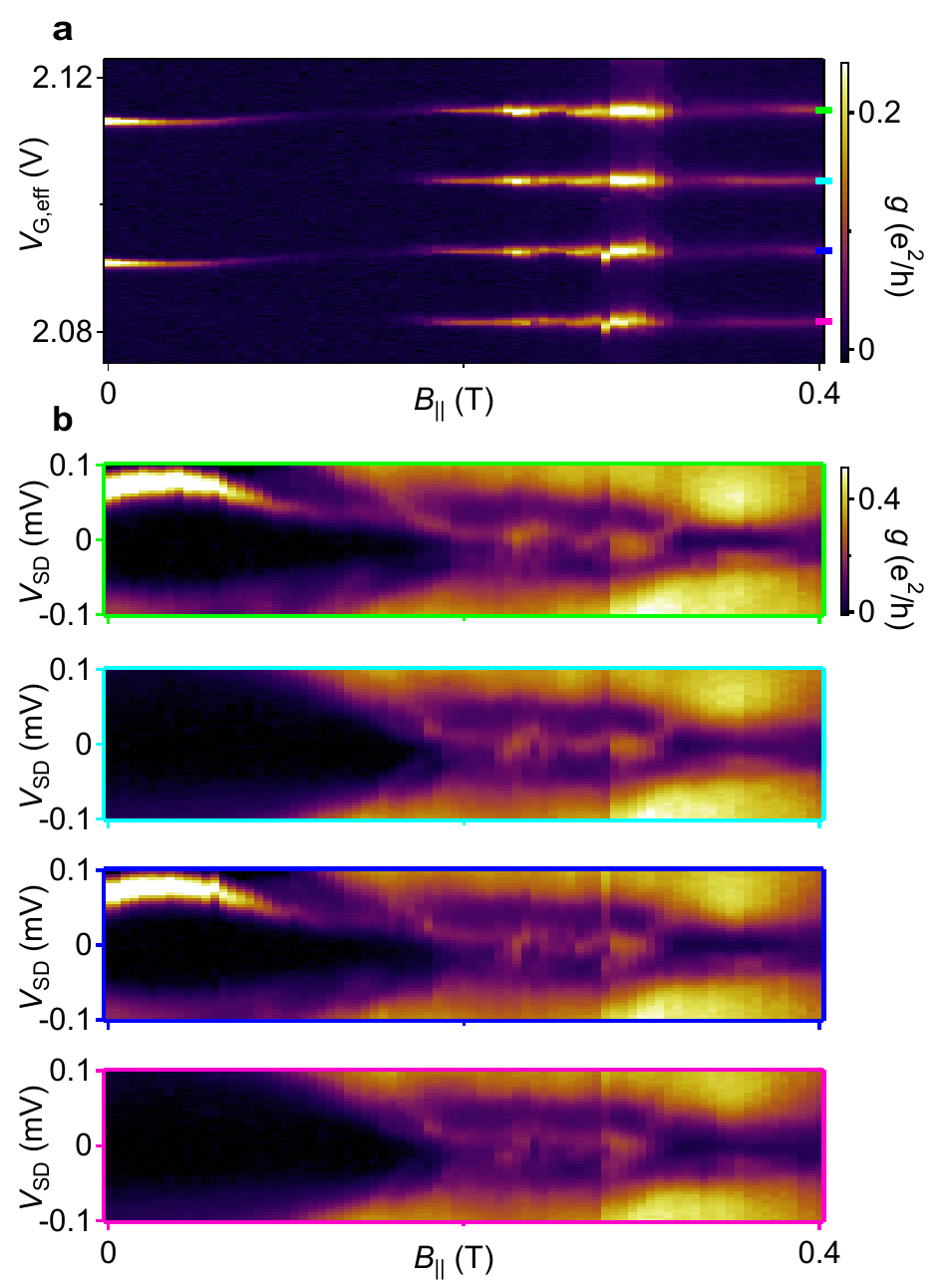}
\caption{\footnotesize{\textbf{Bias-spectroscopy at successive Coulomb peaks.} \textbf{a}, Differential conductance, $g$, versus $V_\mathrm{G,eff}$ and applied magnetic field $B_\mathrm{||}$. $V_\mathrm{G,eff}$  is defined to remove common-mode peak motion, see Methods Section `Bias spectroscopy of long device.'
\textbf{b}, Differential conductance versus source-drain bias, $V_\mathrm{SD}$, and applied magnetic field, $B_\mathrm{||}$, at fixed $V_\mathrm{G,eff}$ indicated on right axis of (a).
}}
\label{fig4zbps}
\end{figure}

\begin{figure}
\center 
\includegraphics{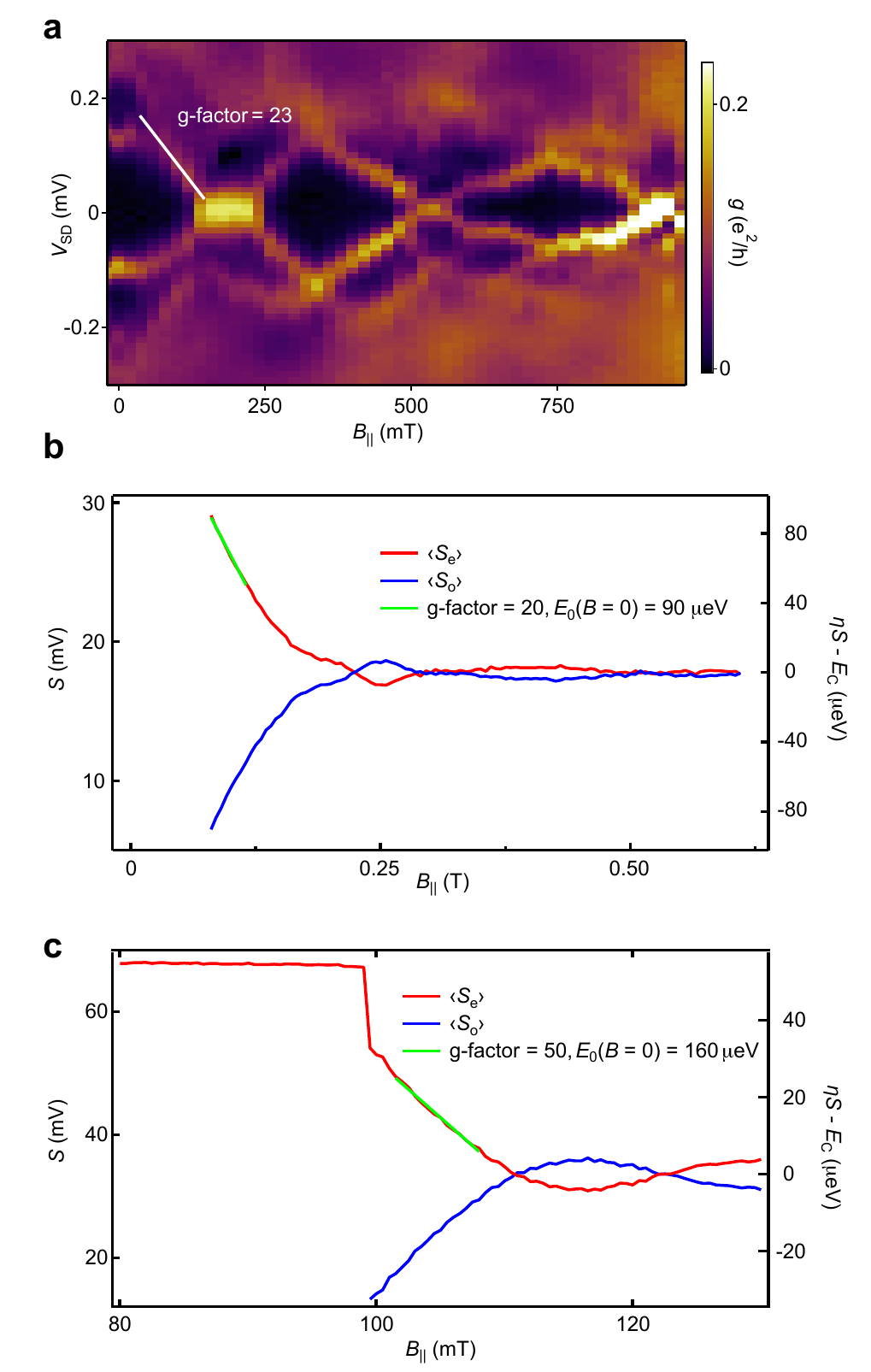}
\caption{\footnotesize{\textbf{Measurement of the g-factor for three devices.} \textbf{a}, Differential conductance, $g$, versus source-drain voltage, $V_\mathrm{SD}$, and applied magnetic field, $B_\mathrm{||}$, for $L = 330~\mathrm{nm}$ showing $\text{g-factor} = 23$.
\textbf{b}, Average even and odd peak spacings $\langle S_ \mathrm{e,o}\rangle $ as a function of $B_{||}$, for $L = 790~ \mathrm{nm} $, showing an extracted $\text{g-factor} = 20$. \textbf{c}, same as in (b) but for $L = 0.9~ \mathrm{\mu m}$ with $\text{g-factor} = 50$.
}}
\label{figgfac}
\end{figure}

\subsection{Measured g-factors}
As can be seen in ED Fig.~\ref{figgfac} the state energy does not move linearly in magnetic field. A non-linear behavior with magnetic field is expected in the presence of strong spin-orbit coupling and a finite critical field.

If the behavior was strictly linear one would expect \begin{equation}
B^{**} = \frac{E_0}{E_0 - E_\mathrm{C}} B^*,
\end{equation}
because the peak splitting at $B^*$ occurs when $E_0(B=0) - E_\mathrm{Z} = E_ \mathrm{C} $ and the state is at zero energy at $B^{**}$ when $E_\mathrm{Z} = E_0(B=0)$ (see Fig. 4 in the main text for reference). The non-linear behavior of $E_0(B)$ at higher magnetic fields approaching $B^{**}$ renders this unsuitable for an accurate measurement of the state energy at zero field.

In the low field regime where the state energy is approximately linear with magnetic field we calculate an effective g-factor. Using this slope it is possible to give a rough estimate of the state energy $E_0(B=0)$ assuming linear behavior and extrapolating the state energy to zero magnetic field.

For bias spectroscopy it should be noted that for gate voltages in the middle of the spectroscopic gap (see main text) transport through a state at $V_ \mathrm{SD} =  V_0$ indicates a state energy $E_0 = e V_ \mathrm{0} /2$. An example for $L = 330~ \mathrm{nm} $ is shown in ED Fig.~\ref{figgfac}a.

Using the addition spectrum, the state energy can be calculated from the peak spacing $S$ according to $E_0 = \left(\eta S - E_ \mathrm{C}\right)/ 2 $. Examples of extracted effective g-factors in the linear range are shown in ED Fig.~\ref{figgfac}b,c.

\end{widetext}

\end{document}